\documentclass{fac}

\usepackage{graphicx}
\usepackage{amssymb}
\usepackage{amsfonts}
\usepackage{amsmath}
\usepackage{dsfont}
\usepackage{MnSymbol}
\usepackage{mathtools}
\usepackage{epsfig}
\usepackage{epstopdf}
\usepackage{tikz}
\usepackage{amsthm}
\ifprodtf \else \usepackage{latexsym}\fi

\usepackage{tikz}
\usetikzlibrary{calc,decorations.pathmorphing,shapes,graphs}

\newcounter{sarrow}

\newcounter{sarrow1}
\newcommand\xnrsquigarrow[1]{%
\stepcounter{sarrow1}%
\mathrel{\begin{tikzpicture}[baseline= {( $ (current bounding box.south) + (0,-0.5ex) $ )}]
\node[inner sep=.5ex] (\thesarrow) {$\scriptstyle #1$};
\path[draw,<-,decorate,
  decoration={zigzag,amplitude=0.7pt,segment length=1.2mm,pre=lineto,pre length=4pt}]
    (\thesarrow1.south east) -- (\thesarrow1.south west);
    $\slashedarrowfill@\relbar\relbar/$
    \end{tikzpicture}}%
}

\makeatletter
\def\slashedarrowfill@#1#2#3#4#5{%
  $\m@th\thickmuskip0mu\medmuskip\thickmuskip\thinmuskip\thickmuskip
   \relax#5#1\mkern-7mu%
   \cleaders\hbox{$#5\mkern-2mu#2\mkern-2mu$}\hfill
   \mathclap{#3}\mathclap{#2}%
   \cleaders\hbox{$#5\mkern-2mu#2\mkern-2mu$}\hfill
   \mkern-7mu#4$%
}
\def\rightslashedarrowfillb@{%
  \slashedarrowfill@\relbar\relbar/\rightarrow}
\newcommand\xnrightarrow[2][]{%
  \ext@arrow 0055{\rightslashedarrowfillb@}{#1}{#2}}

\def\rightslashedarrowfille@{%
  \slashedarrowfill@\relbar\relbar/\twoheadrightarrow}
\newcommand\xntworightarrow[2][]{%
  \ext@arrow 0055{\rightslashedarrowfille@}{#1}{#2}}

\def\rightslashedarrowfillg@{%
  \slashedarrowfill@\relbar\relbar{\raisebox{.12em}{}}\twoheadrightarrow}
\newcommand\xtworightarrow[2][]{%
  \ext@arrow 0055{\rightslashedarrowfillg@}{#1}{#2}}

\def\rightslashedarrowfillx@{%
  \slashedarrowfill@\Relbar\Relbar/\rightrightarrows}
\newcommand\xnTworightarrow[2][]{%
  \ext@arrow 0055{\rightslashedarrowfillx@}{#1}{#2}}

\def\rightslashedarrowfilly@{%
  \slashedarrowfill@\Relbar\Relbar{\raisebox{.12em}{}}\rightrightarrows}
\newcommand\xTworightarrow[2][]{%
  \ext@arrow 0055{\rightslashedarrowfilly@}{#1}{#2}}

\pgfdeclareshape{slash underlined}
{
  \inheritsavedanchors[from=rectangle] 
  \inheritanchorborder[from=rectangle]
  \inheritanchor[from=rectangle]{north}
  \inheritanchor[from=rectangle]{north west}
  \inheritanchor[from=rectangle]{north east}
  \inheritanchor[from=rectangle]{center}
  \inheritanchor[from=rectangle]{west}
  \inheritanchor[from=rectangle]{east}
  \inheritanchor[from=rectangle]{mid}
  \inheritanchor[from=rectangle]{mid west}
  \inheritanchor[from=rectangle]{mid east}
  \inheritanchor[from=rectangle]{base}
  \inheritanchor[from=rectangle]{base west}
  \inheritanchor[from=rectangle]{base east}
  \inheritanchor[from=rectangle]{south}
  \inheritanchor[from=rectangle]{south west}
  \inheritanchor[from=rectangle]{south east}
  \inheritanchorborder[from=rectangle]
  \foregroundpath{
    \southwest \pgf@xa=\pgf@x \pgf@ya=\pgf@y
    \northeast \pgf@xb=\pgf@x \pgf@yb=\pgf@y
    \pgf@xc=\pgf@xa
    \advance\pgf@xc by .5\pgf@xb
    \pgf@yc=\pgf@ya
    \advance\pgf@xc by -1.3pt
    \advance\pgf@yc by -1.8pt
    \pgfpathmoveto{\pgfqpoint{\pgf@xc}{\pgf@yc}}
    \advance\pgf@xc by  2.6pt
    \advance\pgf@yc by  3.6pt
    \pgfpathlineto{\pgfqpoint{\pgf@xc}{\pgf@yc}}
    \pgfpathmoveto{\pgfqpoint{\pgf@xa}{\pgf@ya}}
    \pgfpathlineto{\pgfqpoint{\pgf@xb}{\pgf@ya}}
 }
}
\tikzset{nomorepostaction/.code=\let\tikz@postactions\pgfutil@empty}

\makeatother

\newcommand\black{\ensuremath{\blacktriangleright}}
\newcommand\white{\ensuremath{\vartriangleright}}

  \newcommand\whbl{\white\kern-.1em--\kern-.1em\black}
  \newcommand\blwh{\black\kern-.1em--\kern-.1em\white}
  \newcommand\blbl{\black\kern-.1em--\kern-.1em\black}
  \newcommand\whwh{\white\kern-.1em--\kern-.1em\white}

\newcommand{\sembrack}[1]{[\![#1]\!]}

\newtheorem{theorem}{Theorem}[section]

\newtheorem{proposition}[theorem]{Proposition}

\title[Draft of Fully Abstract Game Semantics for Actors]
      {Fully Abstract Game Semantics for Actors}

\author[Yong Wang]
    {Yong Wang\\
     College of Computer Science and Technology,\\
     Faculty of Information Technology,\\
     Beijing University of Technology, Beijing, China\\
     }

\correspond{Yong Wang, Pingleyuan 100, Chaoyang District, Beijing, China.
            e-mail: wangy@bjut.edu.cn}

\pubyear{2017}
\pagerange{\pageref{firstpage}--\pageref{lastpage}}

\begin{document}
\label{firstpage}

\makecorrespond

\maketitle

\begin{abstract}
Based on the work on the algebraic theory of actors and game semantics for asynchronous $\pi$ calculus, we give the full abstraction proof of game semantics for actors.
\end{abstract}


\section{Introduction}

We ever did some work on game semantics for actors, but failed. Until now, we get the work on the algebraic theory of actors $\textrm{A}\pi$ \cite{AA} and game semantics for asynchronous $\pi$ calculus \cite{GPI}. There are no fresh things under the sun, $\textrm{A}\pi$ is a restriction of asynchronous $\pi$ calculus by adding some type rules.

So, we firstly give the skeleton of full abstraction proof of game semantics for actors, then we detail the proof. And we do not introduce the preliminaries on actors and game semantics for $\pi$, please refer to \cite{AA} and \cite{GPI}.

\section{Proof Skeleton}

The algebra of actors $\textrm{A}\pi$ acts as a restricted asynchronous $\pi$ calculus to ensure the uniqueness, persistence and freshness of actors.

The syntax of the algebra is following.

$$P::=\textbf{0}\quad|\quad \overline{x}\langle \textbf{y},\overline{\textbf{z}}\rangle \quad|\quad x(\textbf{y},\overline{\textbf{z}}).P\quad|\quad !x(\textbf{y},\overline{\textbf{z}}).P\quad|\quad \nu x.P\quad|\quad P\mid Q$$

Then add the type rules of $\textrm{A}\pi$ in \cite{AA} into the the type rules of processes in \cite{GPI}.

Then still use the concepts of arena, justified sequence, $\preceq$, strategy, $\sqsubseteq$, $\odot$, the category of processes $\mathcal{P}$ defined by $\odot$, composition $;$, closed Freyd Category, well opened, the set of interleavings of the justified sequence $\mid$, distributive-closed, trace operator $\mathbf{Tr}$, we can check that the main conclusions in \cite{GPI} still hold: $\mathbf{Tr}$ is a trace operator for $\mathcal{P}$ by adding new interpretation of type rules of $\textrm{A}\pi$.

Finally, by adding the transition rules of $\textrm{A}\pi$ into transition rule of asynchronous $\pi$, we can check that the game semantics for $\textrm{A}\pi$ still is fully abstract:

For any processes $\Gamma\vdash P,Q;\Sigma$, $P\lesssim Q$ if and only if $\sembrack{P}\subseteq\sembrack{Q}$.

\section{Detailed Proof}

(1) Adding the type rules of $\textrm{A}\pi$ in \cite{AA} into the the typing judgements of processes in \cite{GPI}.

We combine the type rules of \cite{AA} and \cite{GPI} together as follows, where $\Gamma$ and $\Sigma$ are the input and output receptionist set of $P$ and $f:\Gamma\rightarrow \Gamma^*$ is a temporary input name mapping function that relates actors in $P$ to the temporary input names they have currently assume, and $f':\Sigma\rightarrow \Sigma^*$ is for output actor names, which are defined in \cite{AA}. We assume that each actor name has a distinct type.

$$\frac{\Gamma,x:T,f\vdash P;\Sigma,f'}{\Gamma,z:T,f\vdash P\{z/x\};\Sigma,f'}$$

$$\frac{\Gamma,f\vdash P;\Sigma,\overline{x}:T,f'}{\Gamma,f\vdash P\{\overline{z}/\overline{x}\};\Sigma,\overline{z}:T,f'}$$

$$\frac{\Gamma,x:S,y:T,\Gamma',f\vdash P;\Sigma,f'}{\Gamma,y:T,x:S,\Gamma',f\vdash P;\Sigma,f'}$$

$$\frac{\Gamma,f\vdash P;\Sigma,\overline{x}:S,\overline{y}:T,\Sigma',f'}{\Gamma,f\vdash P;\Sigma,\overline{y}:T,\overline{x}:S,\Sigma',f'}$$

$$\frac{}{\Gamma,f\vdash\textbf{0};\Sigma,f'}$$

$$\frac{\Gamma,f_1\vdash P;\Sigma,f_1'\quad\Gamma',f_2\vdash Q;\Sigma',f_2'}{\Gamma,\Gamma',f_1\oplus f_2\vdash P|Q;\Sigma,\Sigma',f_1'\oplus f_2'} \textrm{ if }\Gamma\cap\Gamma'=\emptyset\textrm{ and }\Sigma\cap\Sigma'=\emptyset$$

$$\frac{}{\Gamma,\mathbf{y}:\mathbf{S},f\vdash \overline{x}\langle \mathbf{y},\overline{\mathbf{z}}\rangle;\Sigma,\overline{x}:(\mathbf{S},\mathbf{T}), \overline{\mathbf{z}}:\mathbf{T},f'}$$

$$\frac{\Gamma,\mathbf{y}:\mathbf{S},f\vdash P;\Sigma, \overline{\mathbf{z}}:\mathbf{T},f'}{\{x\}\cup\hat{w},x:(\mathbf{S},\mathbf{T}),ch(x,\hat{w})\vdash x(\mathbf{y},\overline{\mathbf{z}}).P;\{\overline{x}\}\cup\hat{w}',ch(\overline{x},\hat{w}')}$$

if $\Gamma-\{x\}=\hat{w};y\notin\Gamma; f=ch(x,\hat{w}),x\in\Gamma;f=ch(\epsilon,\hat{w}),\textrm{otherwise}$ and if $\Sigma-\{\overline{x}\}=\hat{w}';\overline{z}\notin\Sigma; f'=ch(\overline{x},\hat{w}'),\overline{x}\in\Sigma;f'=ch(\epsilon,\hat{w}'),\textrm{otherwise}$.

$$\frac{\Gamma,\mathbf{y}:\mathbf{S},f\vdash P;\Sigma, \overline{\mathbf{z}}:\mathbf{T},f'}{\{x\}\cup\hat{w},x:(\mathbf{S},\mathbf{T}),ch(x,\hat{w})\vdash !x(\mathbf{y},\overline{\mathbf{z}}).P;\{\overline{x}\}\cup\hat{w}',ch(\overline{x},\hat{w}')}$$

if $\Gamma-\{x\}=\hat{w};y\notin\Gamma; f=ch(x,\hat{w}),x\in\Gamma;f=ch(\epsilon,\hat{w}),\textrm{otherwise}$ and if $\Sigma-\{\overline{x}\}=\hat{w}';\overline{z}\notin\Sigma; f'=ch(\overline{x},\hat{w}'),\overline{x}\in\Sigma;f'=ch(\epsilon,\hat{w}'),\textrm{otherwise}$.

$$\frac{\Gamma, x:T, f\vdash P;\Sigma,\overline{x}:T,f'}{\Gamma-\{x\},f|(\Gamma-\{x\})\vdash \nu x.P;\Sigma-\{\overline{x}\},f'|(\Sigma-\{\overline{x}\})}$$

where $ch(\tilde{x})$ is the same to that in \cite{AA}. The reduction semantics, and reduction relation $\twoheadrightarrow$, and structural equivalence are the same to those in \cite{GPI}, and the symbols $P\downarrow$, $P\Downarrow$, $P\lesssim Q$ and $P\simeq Q$ are also the same to those in \cite{GPI}.

(2) Checking if $\mathbf{Tr}$ is a trace operator for $\mathcal{P}$ by adding new interpretation of type rules of $\textrm{A}\pi$.

On game semantics, the concepts arenas $A,B$, moves, enabling relation, initial moves, negative moves, justified sequence, strategies, the least preorder $\preceq$, prefix-closed $\sqsubseteq$, a category of configurations $\mathcal{P}$, disjoint union of forests $\odot$, composition of strategies $;$, restriction $\upharpoonright$, identity strategy on $A$ as $\textbf{id}_A$, so we can get $\mathcal{P}$ is also a well defined category.

By use of the concepts of well-opened, the set of interleavings of the justified sequences $s,t$ as $s|t$, we can define a closed Freyd Category based on $(\mathcal{P},I,\odot)$, which is consisted of $SMC(\mathcal{P},I,\odot)$, a well-defined Cartesian category of abstractions $\mathcal{A}$, and the functor $!$.

So we also can define the trace operator $\mathbf{Tr}$, according to the following rules,
$$\sembrack{\Gamma, y:T, x:S, \Gamma',f\vdash P;\Sigma,f'}=\sembrack{\Gamma,x:S, y:T, \Gamma',f\vdash P;\Sigma,f'};(\textbf{id}_{\sembrack{\Gamma}}\odot \theta_{\sembrack{T},\sembrack{S}}\odot \textbf{id}_{\sembrack{\Gamma'}})$$

$$\sembrack{\Gamma,f\vdash P;\Sigma, \overline{y}:T, \overline{x}:S, \Sigma',f'}=(\textbf{id}_{\sembrack{\Sigma}}\odot \theta_{\sembrack{S},\sembrack{T}}\odot \textbf{id}_{\sembrack{\Sigma'}});\sembrack{\Gamma,f\vdash P;\Sigma,\overline{x}:S, \overline{y}:T, \Sigma',f'}$$

$$\sembrack{\Gamma, z:T,f\vdash P\{z/x\};\Sigma,f'}=\sembrack{\Gamma,x:T,f\vdash P;\Sigma,f'};(\textbf{id}_{\sembrack{\Gamma}}\odot!\triangle_{\sembrack{T}})$$

$$\sembrack{\Gamma,f\vdash P\{\overline{z}/\overline{x} \};\Sigma,\overline{z}:T,f'}=(\textbf{id}_{\sembrack{\Sigma}}\odot!\nabla_{\sembrack{T}});\sembrack{\Gamma,f\vdash P;\Sigma,\overline{x}:T,f'}$$

$$\sembrack{\Gamma,f\vdash\textbf{0};\Sigma,f'} = \bot_{\Sigma,\Gamma}$$

$$\sembrack{\Gamma,\Gamma', f_1\oplus f_2\vdash P|Q;\Sigma,\Sigma',f_1'\oplus f_2'}=\sembrack{\Gamma, f_1\vdash P,\Sigma,f_1'}\odot \sembrack{\Gamma', f_2\vdash Q;\Sigma',f_2'}\textrm{ if }\Gamma\cap\Gamma'=\emptyset\textrm{ and }\Sigma\cap\Sigma'=\emptyset$$

$$\sembrack{\{x\}\cup \hat{w},x:(\mathbf{S},\mathbf{T}),ch(x,\hat{w})\vdash x(\mathbf{y},\overline{\mathbf{z}}).P;\{\overline{x}\}\cup\hat{w}',ch(\overline{x},\hat{w}')} = !\Lambda(\sembrack{\Gamma,\mathbf{y}:\mathbf{S},f\vdash P;\Sigma,\overline{\mathbf{z}}:\mathbf{T},f'});\varrho_{\sembrack{\mathbf{S}},\sembrack{\Gamma} ,\sembrack{\mathbf{T}}}$$ $$\textrm{          };\textbf{der}_{\sembrack{\mathbf{S},\mathbf{T}}}\odot \textbf{id}_{\sembrack{\Gamma}}$$

if $\Gamma-\{x\}=\hat{w};y\notin\Gamma; f=ch(x,\hat{w}),x\in\Gamma;f=ch(\epsilon,\hat{w}),\textrm{otherwise}$ and if $\Sigma-\{\overline{x}\}=\hat{w}';\overline{z}\notin\Sigma; f'=ch(\overline{x},\hat{w}'),\overline{x}\in\Sigma;f'=ch(\epsilon,\hat{w}'),\textrm{otherwise}$.

$$\sembrack{\{x\}\cup \hat{w},x:(\mathbf{S},\mathbf{T}),ch(x,\hat{w})\vdash !x(\mathbf{y},\overline{\mathbf{z}}).P;\{\overline{x}\}\cup\hat{w}',ch(\overline{x},\hat{w}')} = !\Lambda(\sembrack{\Gamma,\mathbf{y}:\mathbf{S},f\vdash P;\Sigma,\overline{\mathbf{z}}:\mathbf{T},f'});\varrho_{\sembrack{\mathbf{S}},\sembrack{\Gamma} ,\sembrack{\mathbf{T}}}$$

if $\Gamma-\{x\}=\hat{w};y\notin\Gamma; f=ch(x,\hat{w}),x\in\Gamma;f=ch(\epsilon,\hat{w}),\textrm{otherwise}$ and if $\Sigma-\{\overline{x}\}=\hat{w}';\overline{z}\notin\Sigma; f'=ch(\overline{x},\hat{w}'),\overline{x}\in\Sigma;f'=ch(\epsilon,\hat{w}'),\textrm{otherwise}$.

$$\sembrack{\Gamma,\mathbf{y}:\mathbf{S}\vdash\overline{x}\langle \mathbf{y},\overline{\mathbf{z}}\rangle;\Sigma,\overline{x}:(\mathbf{S},\mathbf{T}), \overline{\mathbf{z}}:\mathbf{T}}=\bot_{\Sigma,\Gamma}\odot \textbf{app}_{\sembrack{\mathbf{S}},\sembrack{\mathbf{T}}}$$

$$\sembrack{\Gamma-\{x\},f|(\Gamma-\{x\})\vdash\nu x.P;\Sigma-\{\overline{x}\},f'|(\Sigma-\{\overline{x}\})} =\textbf{Tr}^{\sembrack{T}}_{\sembrack{\Sigma-\{\overline{x}\}},\sembrack{\Gamma-\{x\}}}(\sembrack{ \Gamma-\{x\},x:T,f\vdash P;\Sigma-\{\overline{x}\},\overline{x}:T,f'})$$

Then we can get the following two conclusions.

\begin{proposition}
If $M\equiv N$, then $\sembrack{M}=\sembrack{N}$.
\end{proposition}

\begin{proposition}
If $M\rightarrow N$, then $\sembrack{N}\subseteq\sembrack{M}$.
\end{proposition}

(3) Checking if the game semantics for $\textrm{A}\pi$ still is fully abstract.

To show the game semantics is fully abstract with respect to may-equivalence, we will relate the game semantics to the following transition rules.

Then let $\textbf{trace}(P)$ be the set of traces for the configuration $P$. Then we can get that there is also a bijective correspondence between the traces over $(\Gamma,\Sigma)$ and the justified sequences on $\sembrack{\Sigma}^{\bot}\odot\sembrack{\Gamma}$.

Then we get finally get the full abstraction result as follows.

$$x(\mathbf{y},\overline{\mathbf{z}}).P\xrightarrow{x\langle \mathbf{k},\overline{\mathbf{l}}\rangle}P\{\mathbf{k}/\mathbf{y},\overline{\mathbf{l}}/ \overline{\mathbf{z}}\}$$

$$\overline{x}\langle \mathbf{y},\overline{\mathbf{z}}\rangle\xrightarrow{\overline{x}\langle \overline{\mathbf{k}},\mathbf{l}\rangle}[\mathbf{y}\mapsto\overline{\mathbf{k}}]| [\mathbf{l}\mapsto \overline{\mathbf{z}}]$$

$$\frac{P\xrightarrow{\alpha}P'}{P|Q\xrightarrow{\alpha}P'|Q} \quad bn(\alpha)\cap fn(Q)=\emptyset$$

$$\frac{P\xrightarrow{\alpha}P'}{\nu x.P\xrightarrow{\alpha}\nu x.P'}$$

$$\frac{P\xrightarrow{\alpha}Q\quad P\equiv P'}{P'\xrightarrow{\alpha}Q}$$

$$\frac{P\xrightarrow{x\langle \mathbf{k},\overline{\mathbf{l}}\rangle}P'\quad Q\xrightarrow{\overline{x}\langle \overline{\mathbf{k}},\mathbf{l}\rangle}Q'}{\nu x.(P|Q)\xrightarrow{\tau}\nu x.\nu \mathbf{k}.\nu \mathbf{l}.(P'|Q')}$$

\begin{theorem}
For any processes $\Gamma\vdash P,Q;\Sigma$, $P\lesssim Q$ iff $\sembrack{P}\subseteq \sembrack{Q}$.
\end{theorem}

\section{Conclusions}

We give the fully abstract proof for actors, and it is based on the algebraic theory of actors $\textrm{A}\pi$ \cite{AA} and the full abstract game semantics for the asynchronous $\pi$ calculus \cite{GPI}, just because $\textrm{A}\pi$ is a restricted asynchronous $\pi$ calculus.

\newpage

%

\label{lastpage}

\end{document}